\documentclass[nonacm,sigplan,10pt]{acmart}
\renewcommand\footnotetextcopyrightpermission[1]{} 
\acmDOI{}
\acmISBN{}
\acmPrice{}
\usepackage[english]{babel}
\usepackage{times}
\usepackage{blindtext}
\usepackage{tabularx}
\usepackage{graphicx}
\usepackage{dblfloatfix}
\usepackage{color}
\usepackage{xspace}
\usepackage{thumbpdf}
\usepackage{listings}
\usepackage{verbatim}
\usepackage{hyperref}
\usepackage{booktabs}
\usepackage{placeins}
\usepackage{balance}
\usepackage{subfigure}
\usepackage{appendix}
\usepackage{enumitem}
\usepackage{amsmath}
\usepackage{url}
\usepackage{multirow}
\usepackage{multicol}
\usepackage{bm}
\usepackage{algorithm}
\usepackage{algorithmic}
\usepackage{appendix}
\usepackage{soul}

\hypersetup{breaklinks=true}

\newcommand\paragraphb[1]{\noindent{\bf #1.}}
\newcommand\paragraphi[1]{\noindent\emph{#1}}
\newcommand{\bi}{\begin{itemize}}
\newcommand{\ei}{\end{itemize}}

\newcommand{\ie}{{\it i.e.,}\xspace}

\newcommand\eat[1]{}
\newcommand{\allnotes}[1]{}
\renewcommand{\allnotes}[1]{\textit{#1}}

\newcommand{\oursys}{RadioWeaver\xspace}

\hypersetup{pdfstartview=FitH,pdfpagelayout=SinglePage}
\newcommand{\TND}{TND\xspace}
\newcommand{\TNDs}{TNDs\xspace}
\newcommand{\ND}{ND\xspace}
\newcommand{\NDs}{NDs\xspace}
\newcommand{\normalized}{normalized\xspace}

\begin{document}
\pagenumbering{arabic}

\title{5G RAN Slicing with Load Balanced Handovers}


\author{Yongzhou Chen$^1$, Muhammad Taimoor Tariq$^1$, Haitham Hassanieh$^2$, Radhika Mittal$^1$ \\ \emph{$^1$UIUC, $^2$EPFL}}

\maketitle

\noindent{\bf Abstract.} 
With increasing density of small cells in modern multi-cell deployments, a given user can have multiple options for its serving cell. The serving cell for each user must be carefully chosen such that the user achieves reasonably high channel quality from it, and the load on each cell is well balanced.
It is relatively straightforward to reason about this without slicing, where all users can share a global load balancing criteria set by the network operator. In this paper, we identify the unique challenges that arise when balancing load in a multi-cell setting with 5G slicing, where users are grouped into slices, and each slice has its own optimization criteria, resource quota, and demand distributions, making it hard to even define which cells are overloaded vs underloaded. 
We address these challenges through our system, RadioWeaver, that co-designs load balancing with dynamic quota allocation for each slice and each cell. RadioWeaver defines a novel global load balancing criteria across slices, that allows it to easily determine which cells are overloaded despite the fact that different slices optimize for different criteria. Our evaluation, using large-scale trace-driven simulations and a small-scale OpenRAN testbed, show how RadioWeaver achieves 16-365\% better performance when compared to several baselines. 

\sloppy

\section{Introduction}
\label{sec:intro}

\begin{figure}[t!]
\begin{center}
    \includegraphics[width=\linewidth]{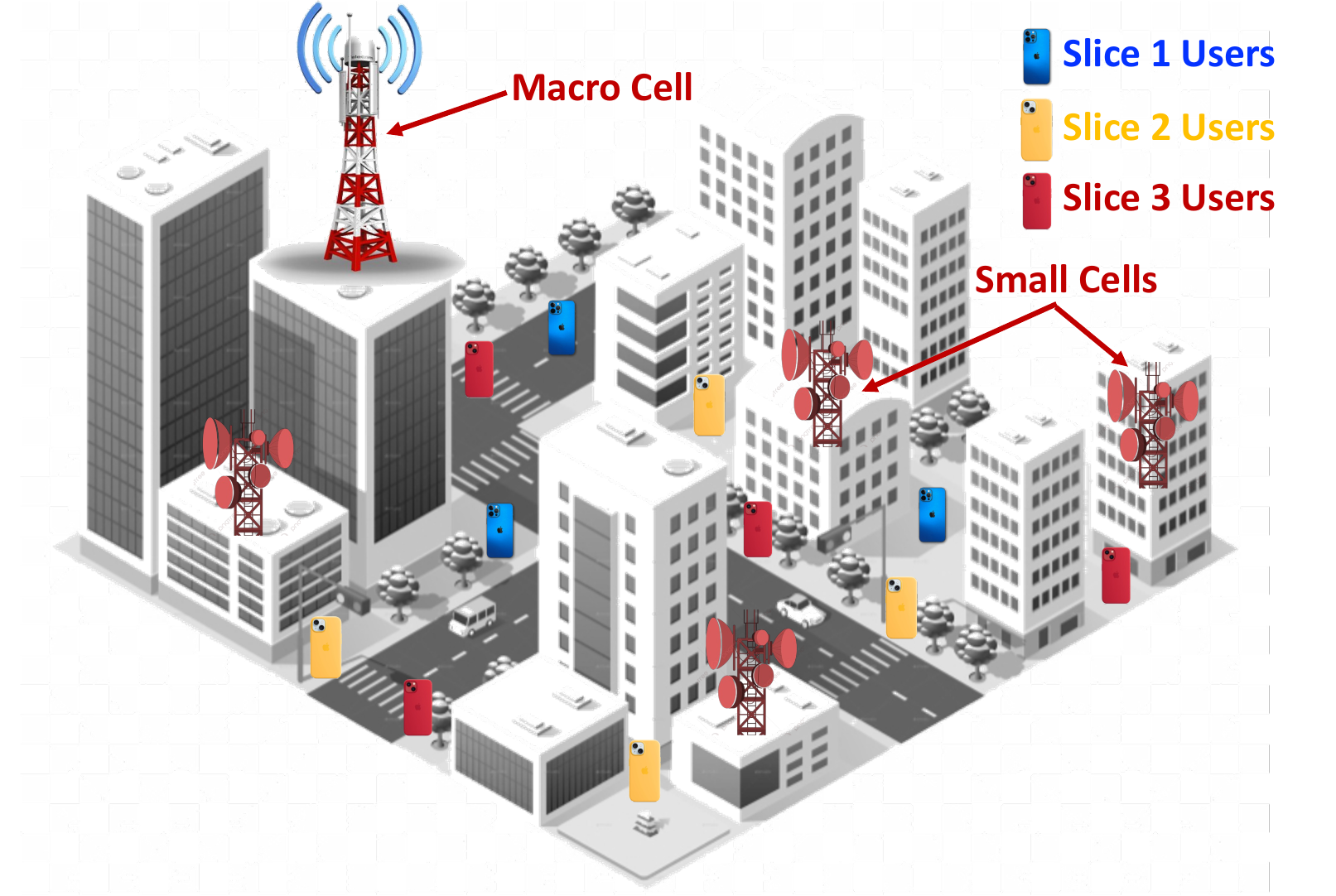}
    \vskip -0.1in
    \caption{Typical multi-cell deployment}
    \vskip -0.25in
\label{fig:MultiCell}
\end{center}
\end{figure}

Recent years have witnessed a sharp increase in the density of cellular base-station deployments, where multiple \emph{small cells} are added in the vicinity of a macro-cell to improve coverage, signal quality, and overall capacity~\cite{netdense, survey-hetero-5g, smallcell-numbers, smallcell-abiresearch}, as shown in Fig.~\ref{fig:MultiCell}. At the end of 2022, there were 142K macro-cell towers across the US, and a total of 452K outdoor small cell nodes~\cite{smallcell-numbers}. It is projected that by 2027, there will be 13 million outdoor 5G small cell deployments at a global level~\cite{smallcell-abiresearch}. 

In such dense deployments, a user can be in the range of multiple cells. To ensure high throughput, as users move, we must choose the serving cell for each user such that the user achieves reasonably good channel quality and the load on each cell is well balanced. Overloading a cell with many users leads to insufficient radio access network (RAN) resources and thereby lowers user throughput. We must therefore handover users from overloaded to underloaded cells without significantly degrading the channel quality of these users. We refer to this problem as \emph{load balanced handovers}. 

Load balanced handovers are particularly challenging in 5G cellular networks due to RAN slicing -- a key 5G feature where the RAN resources are split into large slices owned by different entities (e.g., MVNOs, enterprises, campuses, etc)~\cite{cisco-whitepaper, 3gpp-5gslicing, 5gslice-survey}. The slice owners enter into  service-level agreements (SLAs) with the network providers, that define the type of service for the slice, i.e. its overall quota of resources, number of supported users, and resource scheduling policy (that each slice can customize as per its own optimization criteria, e.g. proportional or datarate fairness, prioritizing certain users, etc.)~\cite{orion, nvs-mobicom,radiosaber,zipper}. 


In the absence of slicing (e.g., in 4G), load-balanced handovers can be easily performed based on the single criterion that the network operator sets as the measure for load (or demand)~\cite{loadbalance-globalpf, loadbalance-smallcell, loadbalance-ml}. For example, as users move, the operator can assign them to the cell with the best channel quality while trying to balance the total number of users across cells or equalize the datarate (throughput) achieved by each user. 
On the other hand, with slicing, each slice has its own optimization criteria that governs its definition of load. While the load for one slice may be defined based on the number of its users at each cell, another slice may define it based on the datarate demands of its users, and yet another slice may weigh users as per their respective priorities when assessing load. 
These differing optimization criteria across slices, combined with their differing quota allocations and user distributions, make it difficult to reason about the overall load on a cell serving users from multiple slices.

A natural solution in such a setting is to perform load-balanced handovers for each slice in isolation from other slices. Each slice gets a statically assigned quota at each cell as per its SLA (e.g. 20\% resources at each cell, for an overall quota of 20\%), and then attempts to independently balance its own set of users across cells as per its own criteria and demand distribution. However, such a solution is overly restrictive and can easily prevent effective load balancing. For example, slice $A$ might have a very high load at cell $C_1$ compared to cell $C_2$, but it might not be possible to handover any of its users from $C_1$ to $C_2$ since they are far from $C_2$'s base station and experience a very poor channel quality at $C_2$. At the same time, another slice $B$ might have many users that experience equally good channel quality at both $C_1$ and $C_2$ (we henceforth refer to such users as ``\emph{amenable}'' users). 
Restricting load balancing to be performed in isolation by each slice results in a sub-optimal outcome in such cases, precluding load balancing for Slice A that has no amenable users. A better outcome could be possible if we can somehow handover the amenable users in Slice $B$ from $C_1$ to $C_2$ in order to free more resources at $C_1$ for slice $A$. But how do we effectively do that without hurting slice $B$? How do we reason about global load balancing of users across different slices that have differing SLAs, optimization criteria, and demand distributions, that must be respected?




In this paper, we introduce \oursys, a system that freely load balances users in multi-cell deployments even when they belong to different slices. \oursys achieves this by co-designing load balancing with flexible resource quota allocation across slices and cells. We vary the quota allocated to each slice at each cell as per their demand distributions while ensuring that the total amount of resources allocated to each slice, when summed across all cells, matches its overall quota. 
For example, 
if slices $A$ and $B$ have an overall quota of 50\% each, we need not divide resources at both cells $C_1$ and $C_2$ in 50-50 ratio for these slices, and can disproportionally allocate say, 25\% and 75\% resources to slices $A$ and $B$ respectively at $C_1$, and conversely 75\% and 25\% resources to slices $A$ and $B$ at $C_2$ (in the case where both cells have equal amount of resources).
Such a disproportional allocation of quota across cells can help meet slice demands. However, it only works if they are \emph{complementary}, i.e. one slice is more loaded at one cell, while another slice is equally more loaded at another cell.


\oursys's load balancing strives to achieve such complementary demand distributions. It does so by defining a novel global load balancing criterion: \TND ({\it total \normalized demand}).
\TND allows \oursys to determine which cells are overloaded despite the fact that different slices have different optimization criteria. To compute \TND, we first compute the relative proportion of radio resources each slice needs at each cell. We can compute this independently for each slice based on its own optimization criteria and user distribution as detailed in \S\ref{sec:design}. We then normalize these per-cell relative demands of each slice with the total network-wide quota of the slice. We finally compute the TND at each cell as the sum of these \normalized demands across all slices at that cell. We use the \TND of a cell as the measure of its global load. A cell is overloaded if its \TND is larger than its capacity. 

\oursys leverages \TND for load balancing as follows. Starting with the current assignment of the serving cell for each user and assuming that displaced and new users are attached to the cells with the highest channel quality, \oursys computes the \TND at each cell to determine whether it is overloaded or underloaded. \oursys's algorithm then iterates by moving amenable users from overloaded to underloaded cells irrespective of what slice they belong to, recomputing the \TNDs that change with user movements. Once the \TND matches the capacity at each cell, we achieve the fully complementary distributions that we seek. \oursys can then allocate the desired \normalized demands to each slice at each cell, allowing each slice to satisfy its optimization criteria. We formally define the conditions for fully complementary distributions and prove how it can maximize the performance for each slice in \S\ref{sec:design}. We also provide an illustrative example of how \oursys works in \S\ref{sec:motivation}.

A few things are worth clarifying about our system: (i) \oursys runs in the centralized RAN controller (OpenRAN's non-realtime RIC in our prototype\cite{scope, colosseum}), and is triggered by significant user movements at time-scales typically ranging from hundreds of milliseconds to a few seconds\cite{handover-frequency}. (ii) \oursys does not trigger any physical handover while the algorithm runs -- it waits for the algorithm to finish and then checks which users' serving cells have changed to trigger a handover for these users.

Realizing \oursys in practical deployment scenarios requires tackling an interesting challenge. The coverage region of overloaded and underloaded small cells need not overlap with one another,  thereby preventing handover of users between small cells for load-balancing. To address this, \oursys takes advantage of the fact that the small cells overlap with the macro-cell. It uses the macro-cell as an intermediate relay. First, we move users from overloaded small cells to the macro-cell, overloading the macro-cell. Then, we move other users from the overloaded macro-cell to the underloaded small cells, repeating this process over multiple rounds as detailed in \S\ref{sec:design}. 

In addition, \oursys must tackle further challenges: (i) How do we compute the \normalized demand for each slice under different optimization criteria like weighted data rate fairness, proportional fairness, or meeting specific datarate demands? (ii) How do we define amenable users? What happens if we run out of amenable users to hand over from overloaded to underloaded cells?  In which order should we move amenable users?  (iii) How do we handle cells with different capacities (e.g. macro-cell vs small cells)? We answer these questions in \S\ref{sec:design} by presenting a complete system design of \oursys. 


We implement and evaluate \oursys using trace-driven simulations and a real world OpenRAN testbed deployed on Colosseum \cite{colosseum}. We compare \oursys to several baselines: 

\noindent (1) Naive load balancing based on the number of users~\cite{loadbalance-smallcell}.

\noindent (2) Flexible quota allocation with handovers based on channel quality without any load balancing (as in NetShare~\cite{netshare}).

\noindent (3) Isolated load balancing by individual slices.

\noindent (4) MORA~\cite{mora}: The closest to our work which assigns serving cells to users in multi-slice settings but is restricted to slices using proportional fairness and adopts an adhoc greedy approach without any explicit load-balancing criteria.

\noindent (5) MORA++: Our enhancement of MORA~\cite{mora} which generalizes it to slices with different objectives by leveraging \oursys's relative demand computation. 

Our results show how \oursys consistently outperforms each of these baselines in terms of slice-level criteria such as fairness, flow completion time of prioritized users, tail throughput, etc. \oursys demonstrates 25-365\%, 35-177\% and 16-105\% better performance than isolated per-slice load-balancing, MORA, and MORA++ respectively, and even stronger wins compared with the other baselines. At the same time, \oursys triggers fewer handovers than other load balancing schemes (1.8$\times$ and 2.4$\times$ lower than MORA and MORA++).



\section{\oursys's Scope}
\label{sec:scope}


In many multi-cell deployments, each cell operates on its own frequency band that does not interfere with others~\cite{5g-diff-bands, 5g-challenges, 5g-heteroband}. Load-balancing across cells (the focus of our work) is a primary factor influencing RAN performance in such scenarios. 
Alternative multi-cell deployments can also allow cells to share the same frequency bands in order to improve spectral efficiency -- such deployments additionally require a more complex RAN resource scheduling system that manages interference~\cite{ericsson-reuse,interference-management, mute-analysis, mute-schedule}. Interference management is beyond the scope of this work. Moreover, ensuring that the network operator can comply with the SLAs of all slices requires an admission control mechanism~\cite{zipper, ml-adm-ctl, slice-overbooking} -- this too is beyond the scope of our work.
We view the above two problems as complementary to \oursys, with admission control taking place before load balancing and interference management (if needed) taking place after.
\oursys's load-balancing, invoked at coarse time-scales of hundreds of milliseconds, is also decoupled from finer-grained radio resource scheduling across users at each cell, done at per-TTI timescales of a few hundred microseconds~\cite{radiosaber}, with both tackling complementary dimensions of optimizing the same slice-specific criteria.
 
\begin{figure*}[t!]
    \centering
    \includegraphics[width=\linewidth]{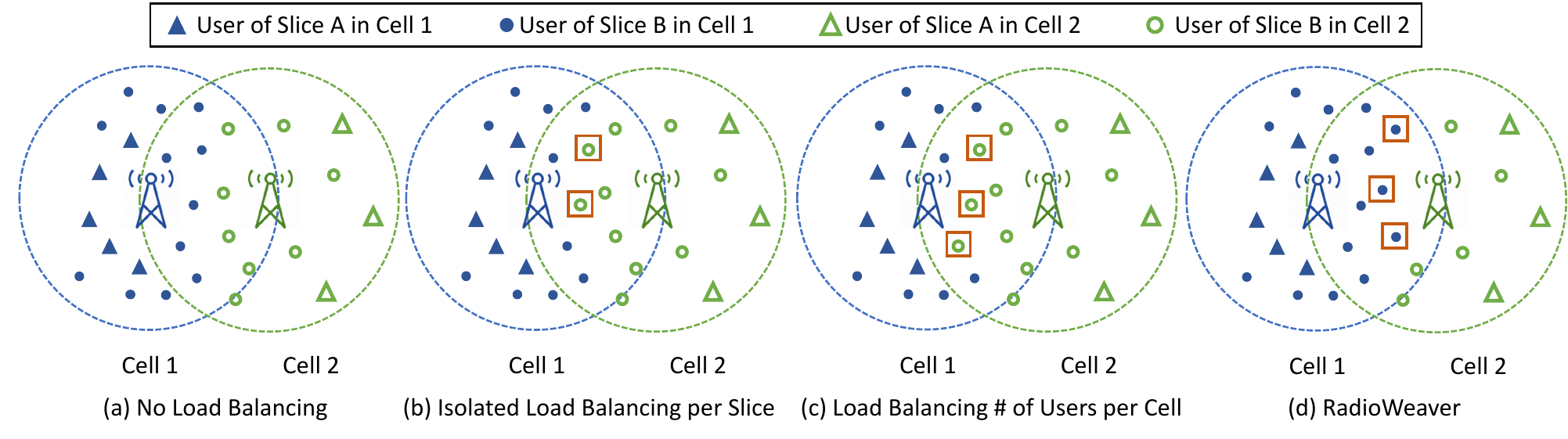}
    \vskip -0.1 in
    \caption{The diagrams represent the coverage areas and user distribution across two cells ($C_1$ represented in blue and $C_2$ represented in green). Users are split among two slices ($S_A$ depicted by triangles and $S_B$ depicted by circles). The color of each user represents the corresponding serving cell under different handover schemes. Red squares highlight the extra handovers caused by load balancing for different schemes.
    }
    \vskip -0.1 in
    \label{fig:motivate1}
\end{figure*}

\begin{figure*}[h]
    \centering
    \includegraphics[width=\linewidth]{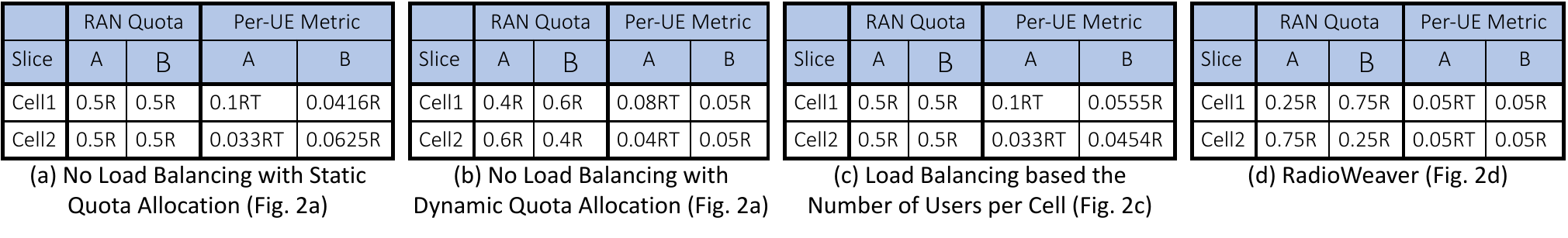}
    \vskip -0.15 in
    \caption{Per-slice RAN quota at each cell and the corresponding per-user metric under different load balanced handover schemes. }
    \vskip -0.15 in
    \label{fig:motivate2}
\end{figure*}

\vskip -0.1 in
\section{Illustrative Examples}
\label{sec:motivation}


We begin with an illustrative example to highlight the challenges in performing load-balanced handovers when there are multiple slices and how \oursys addresses them. Fig~\ref{fig:motivate1} shows 28 users split between two slices $\{S_A, S_B\}$, and distributed in the coverage regions of two cells $\{C_1, C_2\}$. 8 users (denoted as triangles) belong to slice $S_A$, while 20 users (denoted as circles) belong to $S_B$. The user color depicts its serving cell (blue for $C_1$ and hollow green for $C_2$). For simplicity, we assume that both cells have equal capacity ($R$ amount of RAN resources) and both slices have an equal network-wide quota of $R$ each (summed across the two cells). 

Suppose that slice $S_A$ optimizes for data rate fairness, wishing to equalize the throughput across all its users. This requires giving more RAN resources to users with poorer channel quality (who experience a lower data rate per RAN resource block). Slice $S_B$, on the other hand, optimizes for proportional fairness that roughly boils down to equalizing the amount of RAN resources allocated to each user~\cite{loadbalance-globalpf}.  

\vskip 0.06in \paragraphb{(I) No Load Balancing, Only Channel Based Handovers}  We start with considering that each user associates with the cell that gives it the highest channel quality, and there is no load balancing (the status quo today in many deployments\cite{powerho-heterocell}). 
Fig.~\ref{fig:motivate1}(a) depicts the serving cells thus chosen for each user. 
5 users in slice $S_A$ are served by cell $C_1$ while 3 are served by $C_2$. However, these 3 in $C_2$ experience 5$\times$ lower data rate per resource due to poorer channel quality (caused by greater distance from the base-station) than the 5 in $C_1$.
For slice $S_B$, 12 users are served by $C_1$ and 8 users are served by $C_2$.

We first consider the scenario where each slice is allocated equal quota at each cell, i.e. both $S_A$ and $S_B$ get $0.5R$ at $C_1$ and $C_2$. Slice $S_B$ splits its RAN resources equally among its users at each cell. The 12 $S_B$ users served by $C_1$ each get $0.5R/12 = 0.0416R$ resources at $C_1$ and the 8 $S_B$ users served by $C_2$ each get $0.0625R$ resources at $C_2$. Hence, it fails in its objective of proportional fairness among its users.  Slice $S_A$, on the other hand, fairs even worse on its objective of data rate fairness. Suppose each $S_A$ user at $C_1$ experiences a datarate of $T$ per RAN resource. Each $S_A$ user at $C_1$ is allocated $0.1R$ RAN resources and experiences a data rate of $0.1RT$. Each $S_A$ user at $C_2$, in contrast, is allocated $0.167R$ resources, but with 5$\times$ lower channel quality, that results in data rate of $0.167R \times \frac{T}{5} = 0.033RT$. This is 3$\times$ lower than the datarate of $S_A$ users at $C_1$. Hence, it fails in its objective as well.   

\vskip 0.06in \paragraphb{(II) Isolated Handovers} As mentioned in \S\ref{sec:intro}, a simple solution would be to perform load balancing for each slice in isolation of other slices. In this case, slice $S_B$ can easily achieve a more balanced load by migrating two users from $C_1$ to $C_2$ (as depicted in Fig\ref{fig:motivate1}(b)), resulting in 10 $S_B$ users in each cell, with each user getting $0.05R$ RAN resources. On the other hand, slice $S_A$ experiences a higher load at $C_2$ in spite of having fewer users there since these users demand more resources due to their poorer channel quality. However, we cannot move any of $S_A$'s users from $C_2$ to $C_1$ since these 3 users are outside the coverage area of $C_1$. This highlights how isolated handovers can result in sub-optimal outcomes by being overly restrictive, i.e., not every slice may have amenable users that can be easily moved to balance the load.

\vskip 0.06in \paragraphb{(III) Dynamic RAN Quota Allocation} 
Going back to the scenario in Fig~\ref{fig:motivate1}(a), we observe that two slices have somewhat complementary demand distributions: $S_A$ has a higher load at $C_2$ due to the poorer channel quality, while $S_B$ has a higher load at $C_1$ with more users. We can exploit this complementary demand distributions to adjust their relative quotas at the two cells~\cite{netshare}. Specifically, given the 12:8 demand ratio across the two cells for slice $S_B$, we can allocate $S_B$ a quota of $0.6R$ at $C_1$ and $0.4R$ at $C_2$. We can then conversely allocate $S_A$ $0.4R$ at $C_1$ and $0.6R$ at $C_2$, such that both slices get a total of $R$ RAN resources across the two cells as per their SLAs. This results in the desired fairness across all $S_B$ users (each getting $0.05R$ RAN resources across both cells as shown in Fig.~\ref{fig:motivate2}(b)). However, slice $S_A$ is still unable to meet its objective of data rate fairness. Each $S_A$ user at cell $C_1$ now has a datarate of $\frac{0.4R}{5} \times T = 0.08RT$, while each $S_A$ user at cell $C_2$ now has a datarate of $\frac{0.6R}{3} \times \frac{T}{5} = 0.04RT$ (2$\times$ lower than the users at $C_1$). This indicates that while dynamic quota allocation is helpful, it must be co-designed with a global load-balancing mechanism that can make the slice demand distributions fully complementary.

\vskip 0.06in \paragraphb{(IV) Load Balancing Number of Users per Cell} 
To highlight the need for well-designed load-balancing criteria, let us consider what happens under the most natural global load-balancing criteria -- the total number of users attached to each cell\cite{loadbalance-smallcell}. In this case, we start with a total of 17 users (5 $S_A$ users and 12 $S_B$ users) at $C_1$ and 11 users (3 $S_A$ users and 8 $S_B$ users) at $C_2$. This naive criteria would judge cell $C_1$ to be more overloaded than cell $C_2$ and would move three users from $C_1$ to $C_2$. Suppose we move three amenable $S_B$ users, that have similar channel qualities at both cells, from $C_1$ to $C_2$ as shown in Fig\ref{fig:motivate1}(c). The resulting outcome is that $S_B$ now has 9 users at $C_1$ and 11 users at $C_2$ (so a higher load at $C_2$). Slice $S_A$ also continues to have a higher load at $C_2$ (with users having poorer channel quality there). So, instead of complementing the slice demands, the naive criteria ends up making the slice demands fully non-complementary. This results in each slice getting a quota of $0.5R$ RAN resources at each cell and failing to meet its objective as can be seen in Fig\ref{fig:motivate2}(c).

\vskip 0.06in \paragraphb{(V) \oursys} We define a new global load-balancing criteria: \TND (total \normalized demand). \oursys notes that slice $S_B$ has a relative demand ratio of $12R$:$8R$ at the two cells (i.e. 60\% at $C_1$ and 40\% at $C_2$). The \normalized demand (\ND) of $S_B$ at $C_1$ is therefore $0.6 \times (\text{total quota of } S_B) =  0.6R$. Similarly, the \ND of $S_B$ at  $C_2$ is $0.4R$. Slice $S_A$, on the other hand, has a relative demand ratio of $5 \times RT$:$3 \times 5RT =$ 5:15 at the two cells (i.e. 25\% at $C_1$ and 75\% at $C_2$). Therefore, the \ND of $S_A$ is  $0.25R$ at $C_1$ and $0.75R$ at $C_2$. We compute the \TND at each cell by summing up the \NDs of both slices at that cell: $C_1$'s \TND is $0.6R + 0.25R = 0.85R$, while $C_2$'s \TND is $0.4R + 0.75R = 1.15R$. \oursys marks $C_2$ as overloaded since its \TND exceeds its capacity ($R$), and likewise, marks $C_1$ as underloaded. It therefore moves users from $C_2$ to $C_1$ even though $C_1$ has more users to begin with. 

Unlike isolated handovers, there is slice-agnostic flexibility in picking which users are moved. In our example, \oursys moves three amenable $S_B$ users from $C_2$ to $C_1$ (even though $S_B$ has a higher load at $C_1$ to begin with). Now, with 15 users at $C_1$ and 5 users at $C_2$, slice $S_B$'s \ND becomes $0.75R$ at $C_1$ and $0.25R$ at $C_2$. The \TND at each cell now becomes equal to $R$ (i.e. the cell capacities). Therefore, the slice demands are now fully complementary. We can now allocate $S_B$ a quota of $0.75R$ at $C_1$ and $0.25R$ at $C_2$. Conversely, we can allocate $S_A$ a quota of $0.25R$ at $C_1$ and $0.75R$ at $C_2$ as shown in Fig\ref{fig:motivate2}(d). This results in each $S_B$ user getting $0.05R$ RAN resources. Moreover, each $S_A$ user experiences a data rate of $0.05RT$ ($\frac{0.5R}{5} \times T$ at $C_1$ and $\frac{0.75R}{3} \times \frac{T}{5}$ at $C_2$). Thus, the load for each slice is well-balanced across the two cells, and the respective fairness objectives of each slice are met. 

We formalize \oursys's design in \S\ref{sec:design} that naturally generalizes to scenarios with many slices (which may have unequal network-wide quotas) and many cells (which may have unequal capacities). 

Our example assumes that there are some locations where users have high channel quality from multiple cells, and other locations where users experience high channel quality from fewer or just one cell. We used NGScope~\cite{ngscope} to collect channel quality traces from multiple cells at different locations in a campus area -- we find that this assumption indeed holds in practice (results presented in Appendix\S\ref{sec:measure}).






\section{Related Work}
\label{sec:related}



\vskip 0.06in \paragraphb{Load Balancing without Slicing}
Load balancing across multiple base-stations has been explored in the 4G LTE context without slicing ~\cite{loadbalance-rl, loadbalance-smallcell, loadbalance-globalpf}. 
Hasan et al.~\cite{loadbalance-smallcell} considered load balancing under fixed and limited user demands, defining RAN utilization ratio as a measure of cell load, and handing over users from overloaded to underloaded cells based on decreasing channel quality from the target cell.
Ye et al.~\cite{loadbalance-globalpf} investigated the optimal user-to-cell association and quota allocation that maximizes proportional fairness across all users. It proves that the optimal resource allocation for proportional fairness is to equalize the amount of RAN allocated to each user, and proposes a distributed algorithm to figure out the final attachment. Prado et al.~\cite{loadbalance-rl} jointly optimize for proportional fairness and reduction in handover frequency in the multi-cell scenario using deep reinforcement learning.
We focus on the broader problem of load-balanced handovers when users are split across slices with different criteria and demand distributions. SoftRAN\cite{softran} discusses how a centralized software-defined RAN control plane can enable greater optimization opportunities than a decentralized design when balancing load across cells in dense multi-cell deployments -- \oursys indeed leverages such a centralized architecture which has now been standardized through ORAN~\cite{oran}. 
 
\vskip 0.06in \paragraphb{5G RAN Slicing in Multiple Cells}
5G RAN slicing has largely been studied in the context of single base-stations\cite{radiosaber, orion, zipper, nvs-mobicom, scope}. We discuss two relevant works\cite{netshare, mora} that focus on RAN slicing in multi-cell contexts. Netshare\cite{netshare} proposes a framework for dynamically computing demand-driven quota across slices at each cell, while respecting the global network-wide SLAs. However, it assumes the serving cell of each user is fixed and given, and does not consider the complementary problem of load balancing. It frames per-cell quota allocation across slices as a global optimization problem, which it solves using an off-the-shelf solver.
As described in \S\ref{sec:motivation}, \oursys exploits dynamic quota allocation across slices, but it crucially co-designs it with load-balanced handovers that explicitly tries to make the demand distributions across slices fully complementary. Moreover, \oursys's computation of per-cell quotas differs from NetShare's, providing stronger isolation across slices when demand distributions are not fully complementary (as detailed in \S\ref{subsec:swap}). 

MORA~\cite{mora} is the closest related work, that considers the joint problem of resource allocation and user-to-cell association across multiple slices and multiple cells. MORA's resource allocation mechanism restrictively assumes that each slice optimizes for proportional fairness (PF) objective.
MORA greedily inserts each user to the cell that maximizes the user's throughput (as per MORA's resource allocation mechanism). After inserting a user at cell $C$, it iterates through every other user served by $C$ to check whether handing it over to another cell would result in higher throughput, and hands over at most one such user. This triggers a similar handover at the next cell (MORA limits such cascades to three rounds). Our evaluation in \S\ref{sec:eval} shows how the explicit load balancing criteria used by \oursys achieves better performance when compared to MORA's greedy ad hoc approach.
\section{\oursys Design}
\label{sec:design}




\begin{figure}[t]
    \centering
    \includegraphics[width=\linewidth]{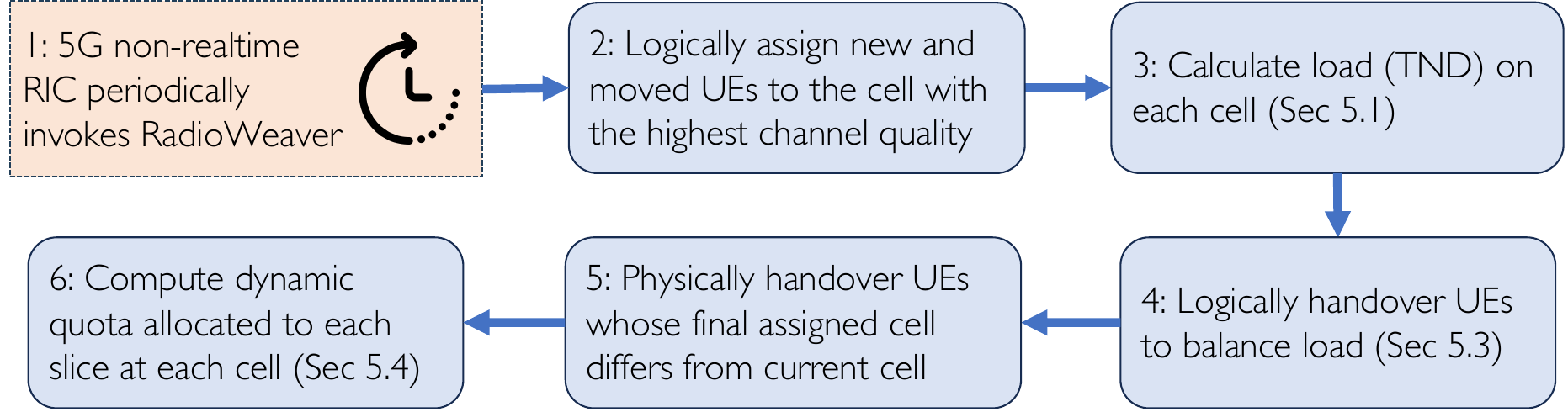}
    \vskip -0.15 in
    \caption{Overall Workflow of \oursys}
    \vskip -0.2 in
    \label{fig:overview}
\end{figure}

\paragraphb{Overview} Fig~\ref{fig:overview} shows the overall workflow of \oursys. (1)  The centralized 5G non-realtime RAN Intelligent Controller (RIC) triggers \oursys whenever it detects significant user movement based on user reports of wideband channel quality indicators (denoted as CQIs)~\cite{cell-reselection-3gpp}. Specifically, our prototype (\S\ref{sec:impl}) triggers \oursys if a user's CQI from its current cell changes by more than a threshold (set to 3dB, same as the standard threshold for triggering handovers\cite{handover_threshold1, handover_threshold2}), checking for this condition every 500ms. (2) Each invocation of \oursys begins with logically assigning each user (referred to as UEs, short for user equipments) the same serving cell as in the previous invocation if its relative location has not changed. For UEs that have moved to a different location or for newly registered UEs, \oursys logically assigns them the serving cell from which they experience highest channel quality. 
(3) \oursys then computes the load on each cell as its \TND (as described in \S\ref{subsec:demand}). 
(4) \oursys then (logically) hands over UEs from overloaded to underloaded cells over multiple rounds, using the macrocell as relay to balance load across non-overlapping small cells (as described in \S\ref{subsec:handover}). It does so iteratively, recomputing the \TNDs that change with the user distribution across cells, until the load is balanced or further handovers are no longer possible. 
(5) Once this process of logical handovers terminates, \oursys checks the final serving cell assigned to each UE, and triggers a physical handover if it differs from the current serving cell of the UE (that was assigned in the previous invocation of \oursys). (6) \oursys finally computes and allocates the dynamic (demand-driven) quotas across slices at each cell (as described in \S\ref{subsec:swap}). 




\subsection{Computing Load on Each Cell}
\label{subsec:demand}

Given a user to cell mapping (initialized at the start of each invocation as described above) \oursys computes the load on each cell as follows: 


\vskip 0.06in \paragraphb{1: Computing demand ratios for each slice} 
\oursys firs computes the fraction of RAN resource blocks (RBs) that a slice wants at each cell (out of its global quota across all cells). We refer to this as the \emph{demand ratio} of the slice, and denote it as $d_{ik}$ for slice $S_i$ at cell $C_k$. By definition, the sum of demand ratio for $S_i$ across all cells ($\sum_{k} d_{ik}$) equals $1$. The demand ratios depend on the slice's optimization criteria (or objective), its user distribution, and the corresponding CQIs. We 
exemplify how to compute the demand ratios for popular objectives.

\vskip 0.06in \paragraphi{(i) Weighted Proportional Fairness (WPF):} The WPF objective for slice $S_i$ corresponds to maximizing the function $\sum_{u_{ij}} w_{ij} log(t_{ij})$, i.e. the weighted sum of log of throughput for each UE $u_{ij}$ in slice $S_i$ (where, $w_{ij}$ denotes weight of $u_{ij}$ and $t_{ij}$ denotes its throughput)~\cite{weighted-pf}. The corresponding demand ratio, $d_{ik}$, for slice $S_i$ at cell $C_k$ is given by $d_{ik} = \frac{\sum_{u_{ij} \in C_k} w_{ij}}{\sum_{u_{ij}} w_{ij}}$, i.e. the weighted sum of UEs in slice $S_i$ that are assigned to cell $C_k$, divided by the weighted sum of all UEs in $S_i$ (across all cells). WPF, thus, reduces to allocating RBs across all UEs in the slice in proportion to their weights. Prior work~\cite{loadbalance-globalpf, mora} shows how this linear relationship with UE weight maximizes the WPF objective. We further validated this both mathematically and empirically (details excluded for brevity).

\vskip 0.06in \paragraphi{(ii) Weighted Datarate Fairness (WDRF):} For a slice $S_i$ optimizing for WDRF, the objective is to ensure that each UE achieves effective throughput (datarate) in proportion to its weight (i.e. $\frac{t_{ij}}{w_{ij}}$ is equalized for all UEs $u_{ij}$ in slice $S_i$). The throughput of $u_{ij}$ (i.e $t_{ij}$) is, in turn, given by $n_{ij} \times e_{ij}$, where $n_{ij}$ is the number of RBs allocated to $u_{ij}$ and $e_{ij}$ is the effective datarate per RB that $u_{ij}$ experiences (as per its CQI from its serving cell). This implies that, given same weights, UEs with poorer channel quality (lower $e_{ij}$) must be allocated more RBs (higher $n_{ij}$). Therefore, the demand ratio of such a slice $S_i$ at cell $C_k$ can be denoted as $d_{ik} = \frac{\sum_{u_{ij} \in C_k} w_{ij} / e_{ij}}{\sum_{u_{ij}} w_{ij} / e_{ij}}$. 


\vskip 0.06in \paragraphi{(iii) Fixed Datarate Demands:} The two forms of fairness above capture saturating demands (e.g. video flows that can adapt their qualities to use up as much datarate as they can get), with weights ($w_{ij}$) capturing user priority. The WDRF formulation can also be applied to slices that specify fixed and limited datarate demands, where the user weights are set in proportion to the demands.



\vskip 0.06in \paragraphb{2: Computing \normalized demand (\ND) of each slice at each cell} Let $Q_i$ denote the global quota of resources (summed across all cells) that must be allocated to slice $S_i$. $Q_i$ is the absolute total quota of $S_i$, computed based on its weight or SLA relative to other slices (as in~\cite{radiosaber, netshare, nvs-mobicom}), such that the sum of quota across all slices ($\sum_i Q_i$) equals the total radio capacity summed across all cells. \oursys computes the \ND for each slice $S_i$ at each cell $C_k$ as $D_{ik} = d_{ik} \times Q_i$ (i.e. the demand ratio of $S_i$ at $C_k$ multiplied by $S_i$'s global quota). By definition, the \ND of each slice, summed across all cells ($\sum_k D_{ik}$) equals the total slice quota $Q_i$.

\vskip 0.06in \paragraphb{3: Computing total \normalized demand (\TND) at each cell} \oursys computes the \TND (aka load) at cell $C_k$ as $L_k = \sum_{i} D_{ik}$ (i.e. the sum of \NDs across all slices at $C_k$). In a way, \TND appropriately \emph{weighs} each user, accounting for both the diverse intra-slice policies and demand distributions, as well as the inter-slice quota allocations. Let the total number of RBs available at cell $C_k$ be $R_k$. 
\oursys marks a cell as overloaded if $L_k > R_k$ and as underloaded if $L_k < R_k$.

\begin{figure}[t]
    \centering
    \includegraphics[width=\linewidth]{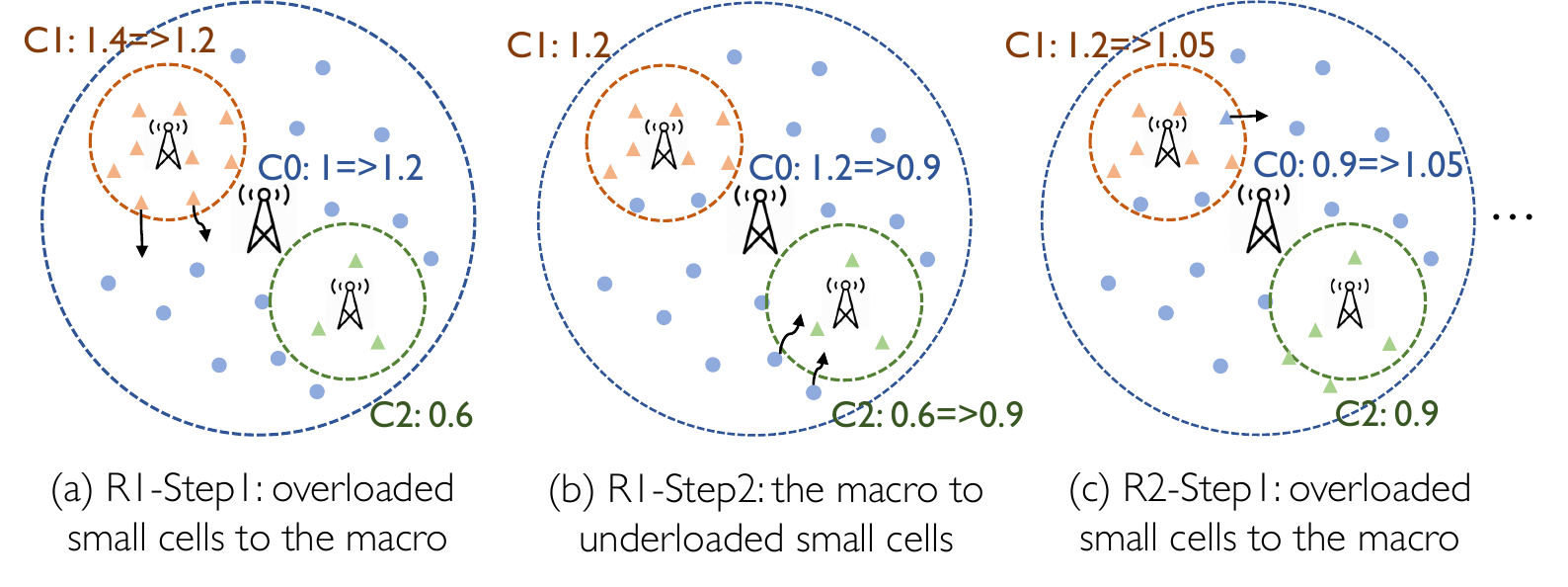}
    \vskip -0.15 in
    \caption{The two-steps handovers of \oursys over multiple rounds. (a)(b) show Step1 and Step2 in Round1, and (c) shows Step1 in Round2.}
    \vskip -0.2 in
    \label{fig:handover}
\end{figure}

\subsection{Co-Designing LB and Quota Allocation}
\label{subsec:theorem}


\oursys co-designs load balancing (LB) with the quota allocation scheme. LB determines the UE to serving cell mapping (we refer to this as the \emph{user distribution}). The allocation scheme determines how the overall quota $Q_i$ of each slice $S_i$ is split across cells. We use the following additional notations: (i) $\mathcal{U}$ for a user distribution. (ii) $\mathcal{Q}$ for an allocation scheme, where $\mathcal{Q}$ is the set $\{Q_{ik}\}$ with $Q_{ik}$ denoting the quota allocated to slice $S_i$ at cell $C_k$. (iii) $P_i(\mathcal{U}, \mathcal{Q})$ denotes the performance achieved by slice $S_i$ on its optimization criteria under a user distribution $\mathcal{U}$ and allocation scheme $\mathcal{Q}$. \oursys tries to find the user distribution and allocation scheme that maximizes the performance of every slice.

\paragraphb{$\bullet$ Lemma 1} For a given user distribution $\mathcal{U}$, $P_i(\mathcal{U}, \mathcal{Q})$ is highest for an allocation scheme $\mathcal{Q}$ where $\forall k, Q_{ik} = D_{ik}$.

\paragraphb{Intuition} In other words, for a given user distribution, the performance of slice $S_i$ is maximized when its allocated quota at each cell is equal to its \normalized demand (\ND) at that cell. This follows from how the demand ratios (and consequently the \NDs) are defined and computed in \S\ref{subsec:demand} (Appendix\S\ref{subsec:proof-lemma1} proves it in the context of WPF and WDRF objectives). 

Let $P^{max}_i(\mathcal{U})$ denote this highest performance achieved by slice $S_i$ under a user distribution $\mathcal{U}$, when it is allocated its desired quota of $D_{ik}$ at each cell $C_k$. 
Allocating the desired quota of $Q_{ik} = D_{ik}$ to every slice at every cell is not possible, as doing so would exceed the available capacity at overloaded cells (with \TND $= \sum_{i}D_{ik} > R_k$). Therefore, it is not always possible for every slice to achieve its maximum performance $P^{max}_i(\mathcal{U})$ under the given user distribution. 

We say that a user distribution $\mathcal{U}$ is \textbf{fully complementary} if the resulting \TND (or load) at each cell is equal to the cell's capacity ($\forall k, L_k = \sum_{i}D_{ik} = R_k$). 

\paragraphb{$\bullet$ Lemma 2} Given a user distribution $\mathcal{U}$, \emph{every} slice $S_i$ can achieve performance $P^{max}_i(\mathcal{U})$ if and only if $\mathcal{U}$ is fully complementary. (Proof in Appendix\S\ref{subsec:proof-lemma2}).



\oursys therefore tries to balance load, moving users from overloaded to underloaded cells, so as to achieve a fully complementary user distribution. 
However, doing so blindly might maximize performance for each slice under the resulting user distribution. But the resulting user distribution itself can be suboptimal if the users get assigned to cells with poor channel quality in the process (that might lower user throughput). We say that a user distribution has \textbf{optimal quality} if each user is attached to one of the cells that maximizes its channel quality. 


\paragraphb{$\bullet$ Theorem 1} A user distribution, $\mathcal{U}^*$, that is both fully complementary and has optimal quality can achieve globally maximum performance $P^{max}_i$ for every slice $S_i$, where $\forall i,  P^{max}_i = \max_{{\mathcal{U}}} P^{max}_i(\mathcal{U})$. (Proof in Appendix\S\ref{subsec:proof-theorem1}).


\oursys's LB therefore strives to achieve a user distribution that is both fully complementary and has optimal quality. However, restricting each user to the cell that maximizes its channel quality might preclude any load balancing for achieving fully complementary distributions. So \oursys relaxes the quality condition by allowing users to be handed off to alternate cells with high enough channel qualities (within a tolerable threshold, $\alpha$, relative to the current cell). This characterizes the set of \emph{amenable} users between each pair of cells. We set $\alpha = 0.8$ in our prototype, and our evaluation in \S\ref{sec:eval} shows how the benefits of load balancing to achieve complementary distributions outweigh the slight compromise in channel quality, improving performance for all slices. 

The ability to achieve fully complementary distribution is restricted by the availability of amenable users (as dictated by users' geographical locations). The next section describes what \oursys does once it runs out of amenable users. 
\oursys's allocation scheme (described later in \S\ref{subsec:swap}) simply assigns the desired per-cell quota to each slice if the user distribution (achieved after \oursys' load balancing) is fully complementary. Otherwise, it assigns per-cell quotas that best meet the desired demands while respecting capacity constraints.

\vskip -0.1 in
\subsection{Logical Handovers for Balancing Load}
\label{subsec:handover}


\oursys's load balancing algorithm logically hands over users from one cell to another over multiple rounds -- we trigger a physical handover only after the algorithm converges and the user distribution has been finalized. Henceforth, we refer to the logical handovers (serving as intermediate steps of our algorithm) simply as handovers. 

\vskip 0.06in \paragraphb{Two-Step Handovers over Multiple Rounds} \oursys balances the load among small cells and macrocell over multiple rounds, with the following two steps in each round: (Step 1) Hand over as many UEs as possible from each overloaded small cell to the macrocell (unless the macrocell becomes relatively more loaded). (Step 2) Hand over as many UEs as possible from the overloaded macrocell to each underloaded small cell (unless the  macrocell becomes relatively less loaded). 
Relieving the overload on the macrocell in Step 2 might open up room for handing over more UEs from overloaded small cells to the macrocell in Step 1 (as illustrated in Fig\ref{fig:handover}). \oursys therefore repeats the round of two steps, and keeps doing so until the load is balanced (i.e. $L_k = R_k$ at all cells) or further handovers are no longer possible (as discussed below).


\vskip 0.06in \paragraphb{Handovers in each step} We now get to which UEs are picked for handovers from an overloaded cell $C_o$ to an underloaded cell $C_u$ in each step (within each round) above. \oursys sorts UEs currently assigned to $C_o$ by the ratios of their channel quality (more specifically, datarate per RB) from the target cell $C_u$ over the channel quality from the source cell $C_o$. It then attempts to hand over UEs from $C_o$ to $C_u$ in that order over two phases (slice-agnostic Phase 1 followed by slice-specific Phase 2), stopping when either of the two conditions are met: (i) Load-driven criteria, triggered when the load on $C_o$ and $C_u$ become equal. As mentioned above, further handovers may still occur in subsequent rounds until the network-wide load is balanced. (ii) Channel-driven criteria that capture the impact of channel quality on performance and differ between the two phases as described below.

\vskip 0.06in \paragraphb{Phase 1 (slice-agnostic): Hand over amenable users} 
\oursys labels a UE as amenable for handovers from $C_o$ to $C_u$ if the ratio of its datarate per RB from $C_u$ to that from $C_o$ is higher than a threshold ($\alpha$) -- such UEs are at the head of the queue that is sorted by this ratio as mentioned above. \oursys hands over such UEs from $C_o$ to $C_u$ (agnostic of the slices they belong to), until it hits the load-driven stopping criteria mentioned above.
As and when each UE is handed over from $C_o$ to $C_u$, the demand ratios for the corresponding slice (and the \NDs at each cell) changes. \oursys recomputes the \TNDs (loads) by repeating the steps in \S\ref{subsec:demand}. 
\oursys ends the slice-agnostic Phase 1 once it runs out of amenable UEs.

\vskip 0.06in \paragraphb{Phase 2 (slice-specific): Hand over users if it helps their slices}
If we run out of amenable UEs, but the load-driven stopping criteria has not yet been met, \oursys transitions to its next phase for handovers. Here, it checks if handing over the next UE $u_x$ in the sorted queue would improve the corresponding slice's objective.
If yes, it proceeds with the handover. Otherwise it skips that UE and checks for the next. The rationale here is that while the handover of $u_x$ from $C_o$ to $C_u$ comes at the cost of higher degradation in channel quality, it can still be allowed if the overall benefits of greater balance in load between $C_u$ and $C_o$ for the slice outweigh the penalty of channel degradation. The phase 2 handovers (and therefore the handovers in that step) stop when either the load-driven criteria is met, or handover of any remaining (non-amenable) UE at $C_o$ would degrade the corresponding slice's objective. 





\vskip 0.06in
While our handover design has been explained under the common context where small cells (with limited range) do not overlap with each other, requiring a relay via the macrocell for load balancing, it can be easily adapted for scenarios where multiple (small) cells overlap with one another (as detailed in Appendix\S\ref{subsec:overlap-smallcells}).

\subsection{Swap-Based Dynamic RAN Allocation}
\label{subsec:swap}

As mentioned in \S\ref{subsec:handover}, the rounds of handovers halt under either of the two conditions: (i) if the load is balanced (no slice is overloaded or underloaded), or (ii) if further handovers between overloaded and underloaded cells are not possible because they would cause unacceptable performance degradation. The first case implies that the total normalized load is equal to the capacity at each cell $C_k$ (i.e. $L_k = \sum_{i} D_{ik} = R_k$). The slice demands are therefore fully complementary -- the resource quota we allocate to each slice $S_i$ at each cell $C_k$ can simply be equal to its \ND ($D_{ik}$) at the cell.

However, if handovers stop due to the second case, it implies that the slice demands are not fully complementary. There is at least one cell $C_{k'}$ where $L_{k'} > R_{k'}$, and where the normalized demand of each slice cannot be met. In such a case, we need to allocate the per-cell quota across slices in a manner that adheres to the partially complementary demands to the best possible extent.

We first tried to apply NetShare's mechanism~\cite{netshare} to compute quota allocations in this case. However, we found that NetShare's global utility function does not guarantee performance isolation across slices. Suppose two slices with equal weights, $S_A$ and $S_B$, both have higher demand ratios at cell $C_1$ and lower at $C_2$, but $S_A$'s demand differential is higher, NetShare may end up allocating a higher quota of resources to $S_A$ compared to $S_B$ at $C_1$, thereby favoring $S_A$ and hurting $S_B$. Specifically, $S_B$ would have fared better under static allocation of equal quota to both slices at each cell.    

\oursys, instead, employs an intuitive swap based algorithm to dynamically compute quotas of each slice at each cell, that retains performance isolation across slices (ensuring, by design, that no slice would have been better off with the static quota allocation at each cell).



\oursys initializes $Q_{ik}$ (quota allocated to slice $S_i$ at cell $C_k$) as per its static share by dividing the global slice quota $Q_i$ across cells in proportion to their capacities. \oursys then greedily picks a pair of slices, say $S_A$ and $S_B$, which have complementary demand ratios in two cells, say $C_1$ and $C_2$. This can occur when $S_A$'s quota allocation is higher than its demand at $C_1$ and lower at $C_2$, while it is the other way round for $S_B$ (i.e. $Q_{A1} > D_{A1}$, $Q_{A2} < D_{A2}$, $Q_{B1} < D_{B1}$, and $Q_{B2} > D_{B2}$). 
\oursys computes the swappable RAN amount $\delta$ to be $min(Q_{A1} - D_{A1}, D_{A2} - Q_{A2}, D_{B1} - Q_{B1}, Q_{B2} - D_{B2})$. \oursys then decreases $Q_{A1}$ and $Q_{B2}$ by $\delta$ and increases $Q_{B1}$ and $Q_{A2}$ by $\delta$ (i.e. it reallocates $\delta$ RBs from $S_A$ to $S_B$ at $C_1$, and conversely from $S_B$ to $S_A$ at $C_2$, keeping the total global quota of each slice unchanged). 
\oursys keeps making such swaps greedily, until it can no longer find any pair of slices with complementary demands in any two cells. 


\subsection{Slice Interface}
\label{subsec:sliceinterface}

The load-balanced handover decisions must be made by the network operator, leveraging the global view of user distribution across slices.
Similar to what was proposed in prior work~\cite{radiosaber}, individual slices can use expressive parameters to express their desired optimization criteria (that governs their demand ratios). 
Specifically, each UE $u_{ij}$ in slice $S_i$ can be assigned a parameterized weight of $w_{ij}/e_{ij}^{(1 - {\epsilon}_i)}$, where the slice specifies weights $w_{ij}$ (updated dynamically as per user priority or relative demands), and the ${\epsilon}_i$ parameter that determines how much to weigh in channel quality ($\epsilon_i = 1$ for proportional fairness, $\epsilon_i = 0$ for datarate fairness, etc). These parameters can also be directly derived from the related parameters that the slice may orthogonally specify for resource scheduling based on the same optimization criteria~\cite{radiosaber}. 

\section{Implementation}
\label{sec:impl}


We evaluate \oursys using large-scale trace-driven simulations (as described in \S\ref{sec:eval}) and smaller scale experiments on a real system deployed in the Colosseum testbed~\cite{colosseum}. The implementation on Colosseum is done in SCOPE~\cite{scope}. SCOPE is a softwarized cellular open prototyping environment, which comprises of 3GPP and OpenRAN compliant base stations (running production-grade srsRAN~\cite{srsran} software) and UEs. SCOPE supports multiple base stations and RAN slicing in the frequency domain.
We implement \oursys module as part of the non-realtime RAN Intelligent Controller (RIC) in SCOPE.
\oursys uses the APIs provided in SCOPE to collect relevant user information (slice id, CQIs, etc). It then runs its load balanced handover logic (as described in \S\ref{sec:design}) to calculate the user-to-cell association and the quota allocation for each slices in each cell. It then enforces these decisions via SCOPE's APIs. 
Our changes to SCOPE fit within 600 lines of code. \oursys' computation of user-to-cell mapping and per-cell quota in each invocation fits within 10 milliseconds (adding only 14\% overhead over typical handover times of 50-90 milliseconds\cite{handover-time1, handover-time2}).



\section{Evaluation}
\label{sec:eval}


Our evaluation  results in the following key takeaways:


\noindent $\bullet$ \oursys is able to effectively balance load to achieve fully complementary distribution across slices and cells (\S\ref{subsec:case1}).

\noindent $\bullet$ This allows \oursys to achieve superior performance on individual slice objectives compared to other baselines (\S\ref{subsec:case1}). 

\noindent $\bullet$ \oursys achieves this while triggering fewer handovers compared to other baselines (\S\ref{subsec:case1}). 

\noindent $\bullet$  Our key trends hold across multiple scenarios as we vary how users are split across slices and cells (\S\ref{subsec:vary}).

\noindent $\bullet$ \oursys's performance wins are also seen in the smaller scale evaluation of our real system implementation on Colosseum testbed (\S\ref{subsec:colosseum}).





\subsection{Baselines}
\begin{figure*}[t!]
    \centering
    \includegraphics[width=\textwidth]{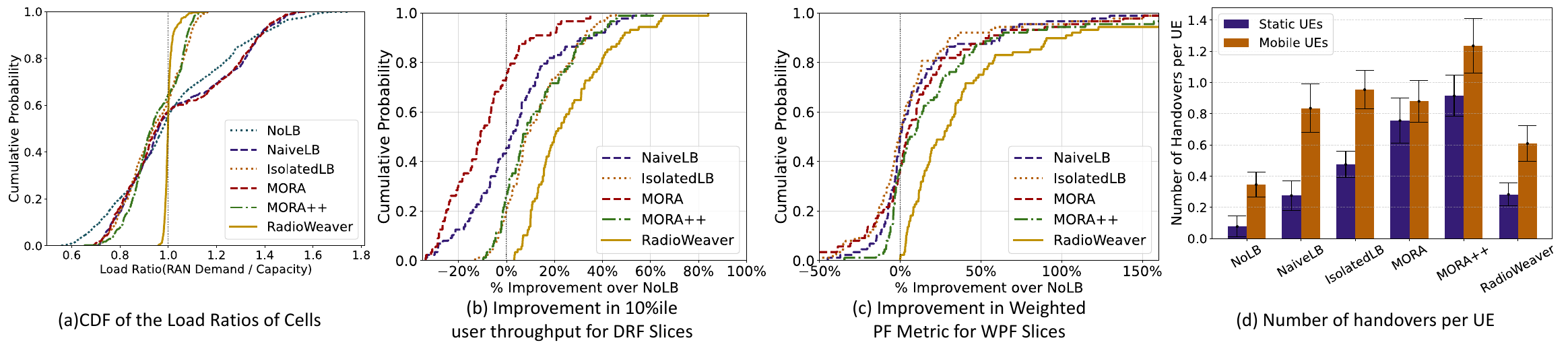}
    \vskip -0.15 in
    \caption{Comparing \oursys and baselines: (a) The CDF graph of cells' load ratios. The load balancing effect on slice-level performance when compared to NoLB in (b) DRF slices and (C) WPF slices. (d) The number of physical handovers triggered per user across the experiment.}
    \label{fig:case1}
    \vskip -0.1 in
\end{figure*}

We compare \oursys with the following baselines 

\vskip 0.04in \paragraphb{(1) No load balancing: CQI-based handovers (NoLB)} This represents the status quo strategy to connect each UE to the cell from which it has highest channel quality without any attempt to balance the load. We use \oursys's dynamic resource allocation mechanism for this baseline, that tries to exploit partially complementary distribution across slices, as described in \S\ref{subsec:swap} -- this performs strictly better than the static quota allocation across slices at each cell. 


\vskip 0.04in \paragraphb{(2) Naive load-balanced handovers 
 (NaiveLB)~\cite{loadbalance-smallcell}} This is the most natural load balancing baseline that uses the total number of users attached to a cell as the proxy for load. At a high-level, NaiveLB uses the same framework as \oursys, but differs in its definition of load -- moving amenable UEs from overloaded cells (with more users) to underloaded cells (with fewer users). The load computed for the macro cell, that has higher RAN capacity than small cells, is proportionally reduced when doing the comparison.  
 
 



\vskip 0.04in \paragraphb{(3) Isolated intra-slice handover (IsolatedLB)} In this baseline, each slice is allocated its static quota at each cell and individually balances its own load. We use \oursys' two-phase logic for this baseline. We first attempt to move the amenable users of the given slice from overloaded cells (where the relative demand of the slice exceeds its assigned quota) to underloaded cells, and then attempt to move other non-amenable users as long as it improves the slice objective. 

\vskip 0.04in \paragraphb{(4) MORA\cite{mora}} MORA assumes all slices optimize for proportional fairness when computing the per-cell quota of each slice. It determines user-cell attachment using a greedy approach (without any explicit load balancing criteria) as described in \S\ref{sec:related}. 

\vskip 0.04in \paragraphb{(5) MORA++} We implement MORA++ as a variant of MORA where we replace its resource allocation logic (specific to all slices optimizing for proportional fairness), with the \oursys's more generalized resource allocation that supports different objectives across slices (using the same greedy user-cell attachment algorithm as MORA)

\subsection{Experimental Settings}
\label{subsec:exp-setting}

We implement \oursys in an open-source RAN simulator~\cite{radiosaber, lte-simulator} fed with real-world traces collected using NGScope~\cite{ngscope}.
We simulate multi-cell scenarios with one macro-cell and varying number of small cells (four in our default scenario). 
We configure the macro-cell with a transmission power of 49 dBM and with 100Mhz radio channel (characteristic of 5G NR\cite{3gpp-400mhz}). 
We configure each small cell with a lower transmission power of 35 dBM and 20MHz radio channel (on a different non-interfering band with respect to the macro-cell). 
We position each small cell at a randomly chosen distance between 500-700 meters from the macrocell, while ensuring the small cells are non-overlapping and spaced at least 1000 meters apart from one another (thus representing realistic multi-cell deployment~\cite{quickc, 3gpp-comp-eval-parameters}). We also evaluate a scenario with overlapping small cells in \S\ref{subsec:overlap-smallcells} to show the generality of \oursys' high-level design.

Our experiments involve 280-350 UEs split across 8 slices and spatially distributed in the multi-cell region as described in the context of individual experiments. At the start of the simulation, each UE is located as per the configured spatial distribution, but half users move over time in random directions at a speed of 18 miles per hour.
We use the simulator's in-built 3GPP urban area path loss model to determine the RSRP a UE would experience from different cells based on the distance. We then assign each UE a real-world trace (collected using NGScope) with per-RB CQIs, such that the average RSRP of the trace matches the modeled value. 


\subsection{Results}
\label{subsec:case1}
\vskip 0.06in \paragraphb{Setup} We first evaluate a scenario where we have 8 slices ($S_1$...$S_8$) with equal global quotas (12.5\% each).
We configure slices with odd index ($S_1$, $S_3$, etc) to have randomly chosen 8-12 users each within range of small cells $C_1$ and $C_2$, and 0-4 users each in the range of small cells $C_3$ and $C_4$.  In contrast, slices with even index ($S_2$, $S_4$, etc) have 0-4 users each in the range small cells $C_1$ and $C_2$, and 8-12 users each in the range of small cells $C_3$ and $C_4$. In addition, each slices has a random number of users (between 5-30) in the macro cell ($C_0$) region, outside the range of small cells. This is depicted at Scenario 1 in Fig\ref{fig:multi-scenarios}(a) (we present results with other scenarios in \S\ref{subsec:vary}). Every user's geo-location is uniformly and randomly generated in the coverage area of the assigned cell, with each user moving over time in random directions. We run our experiments 25 times, with different random seeds that changes the specific number of users, their initial geo-location, and moving directions.
The first four slices ($S_1$ to $S_4$) optimize for weighted proportional-fairness (we refer to these as WPF slices). The last four slices ($S_5$ to $S_8$) optimize for datarate-fairness (referred as DRF slices). We randomly select half of the users from WPF slices in $C_1$ and $C_3$ to be prioritized with 5$\times$ higher weights. Every user runs a single backlogged flow (e.g. representing a video flow for AR/VR, gaming, HD streaming, etc with saturating demands). 

\vskip 0.06in  \paragraphb{1. Load Ratio Across Cells} We define a cell's load ratio to be its TND divided by its RAN capacity, and use it as a measure of the effectiveness of load balancing. We would like the load ratio to be close to 1 for all cells -- a cell is overloaded if the load ratio is greater than 1, and underloaded if it is smaller than 1. We record the load ratio of each cell after each invocation of \oursys for each experiment run. Fig~\ref{fig:case1}(a) shows the corresponding CDF. 
\oursys, which explicitly uses the load ratio as its load balancing criteria, always achieves a load ratio close to 1, deviating in some tail cases where the geo-distribution of users is such that further load balancing would degrade overall performance by degrading channel qualities (as discussed in \S\ref{sec:design}). 
In comparison, other baselines have load ratios further away from 1 in most cases --  NoLB does not even attempt to balance load; NaiveLB uses a different criteria that does not capture slice-specific demands; MORA and MORA++ use a greedy heuristic without any explicit load balancing criteria; and IsolatedLB is too restrictive. \oursys's superior balance in load translates to improved slice-level performance as discussed next.


\vskip 0.06in \paragraphb{2: Performance of DRF Slices} We measure the performance of DRF slices in terms of the 10\%ile throughput computed across users within each such slice. It captures how higher fairness implies improved performance for tail users, and unlike other fairness metric (e.g. max-to-min ratio or Jain's fairness index) it also captures the absolute throughput.   
For every slice, we do a slice-by-slice comparison of each scheme against NoLB as the reference baseline (i.e. the improvement in each slice's metric compared to NoLB).
Fig.~\ref{fig:case1}(b) presents the CDF graph of this improvement across 100 such slices (4 DRF slices per experiment across 25 runs).
For more than half slices, \oursys improves the 10\%ile throughput by 23\% to 84\% compared with NoLB. \oursys also outperforms all baselines. In the median case, \oursys's performance is 21\% and 38\% higher than NaiveLB and MORA respectively (that are both misaligned with slice-specific demands) -- in fact 50\% of DRF slices experience \emph{worse} performance with MORA and NaiveLB when compared to NoLB. \oursys's median performance on this metric is also 12\% higher than IsolatedLB that is too restrictive, and 13\% higher than greedy MORA++.



\vskip 0.06in \paragraphb{3: Performance of WPF Slices} For the WPF slices, we consider each slice's weighted PF objective, i.e. sum of the weighted logarithm throughput of each user in the slice. Fig\ref{fig:case1}(c) shows the CDF of improvement in this WPF metric across each such slice over 25 runs, for each scheme when compared to NoLB.
\oursys is the only system that improves the weighted PF metric in \emph{all} slices compared with NoLB. For 50\% slices, the performance improvement ranges from 31\% to 150\%. MORA++ has the second best result -- it improves the weighted PF metric of 60 out of 100 slices. In the median case, \oursys's improvement is (i) 16\% higher than MORA++, (ii) 20\% higher than MORA, (iii) 27\% higher than NaiveLB, and (iv) 25\% higher than IsolatedLB. 




\vskip 0.06in \paragraphb{4: Number of Handovers}
In Fig~\ref{fig:case1}(d), we compare the number of physical handovers triggered per user (split between stationary and mobile users) for 20-seconds experiment run, averaged across all experiment runs (with the bars showing standard deviations). Compared with \oursys, NaiveLB has $1.37\times$ more handovers for mobile users and similar handovers for static users. IsolatedLB incurs $1.68\times$ more handovers for static users and $1.57\times$ more handover for mobile users than \oursys. The total numbers of handovers of MORA and MORA++ are the highest, owing to their greedy user-cell assignment that creates a cascade of handovers. 
MORA++, which is closest to \oursys in terms of slice-level performance, incurs $2\times$ more handovers for mobile users and $3.23\times$ more handovers for static users than \oursys. All load balancing schemes (including \oursys) naturally incur more handovers than NoLB (that simply attaches each user to the cell with highest channel quality), including more handovers for stationary users that might need to be handed over in order to balance load as the cell assignment of a mobile user changes. 

\vskip 0.06in \paragraphb{5: Varying workload and performance metric} We repeat the above experiments, but with a different workload. Specifically, for slices $S_1$ to $S_4$ that optimize for WPF, we configure the prioritized users, that have 5$\times$ higher weights, to generate web flows following a heavy-tail distribution with Poisson inter-arrival times, with an average data rate of 3Mbps. 
The non-prioritized users in these slices run backlogged flows. We measure the improvement in median flow completion time (FCT) that each scheme achieves over NoLB for the prioritized users in these slices in each experiment run.  Fig\ref{fig:case3}(a) shows the results (when averaged across 25 runs).
\oursys achieves 5.5 $\times$ lower median FCT than NoLB due to its demand-aware load balancing, and 5.1$\times$ lower than NaiveLB by better catering to increased weights of high priority users. \oursys also achieves 3.5$\times$ lower median FCT than IsolatedLB that has restricted load balancing capabilities. \oursys, with its systematic load balancing, achieves 6.7$\times$ and 2.5$\times$ lower median FCT than greedy MORA and MORA++. 

\subsection{Varying Experimental Setup}
\label{subsec:vary}

\begin{figure}[t!]
    \centering
    \includegraphics[width=0.5\textwidth]{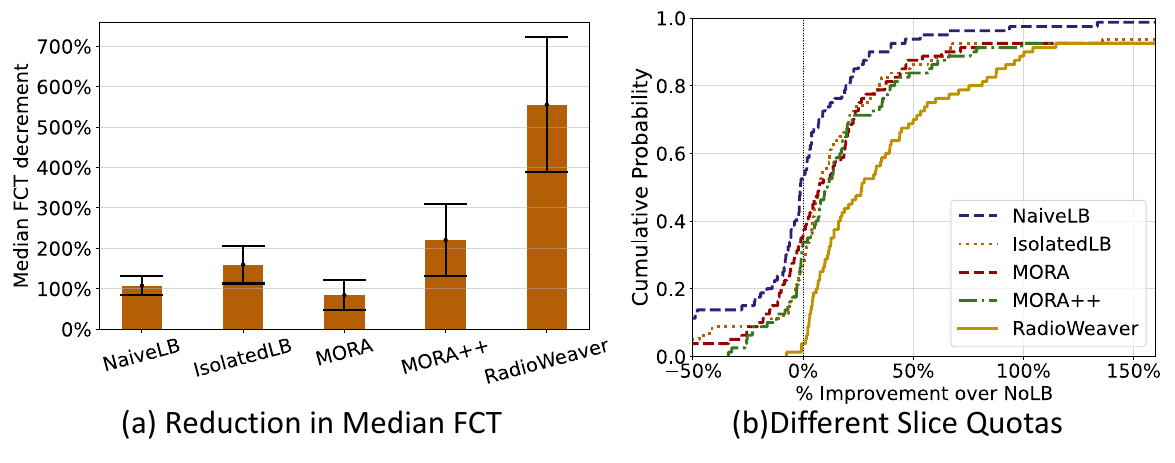}
    \vskip -0.1 in
    \caption{Comparing \oursys and baselines: (a) The reduction in median FCT of web flows. (b) The weighted PF metric improvement of WPF slices with different slice quotas in Scenario3.}
    \label{fig:case3}
    \vskip -0.15 in
\end{figure}

\begin{figure}[t!]
    \centering
    \includegraphics[width=0.5\textwidth]{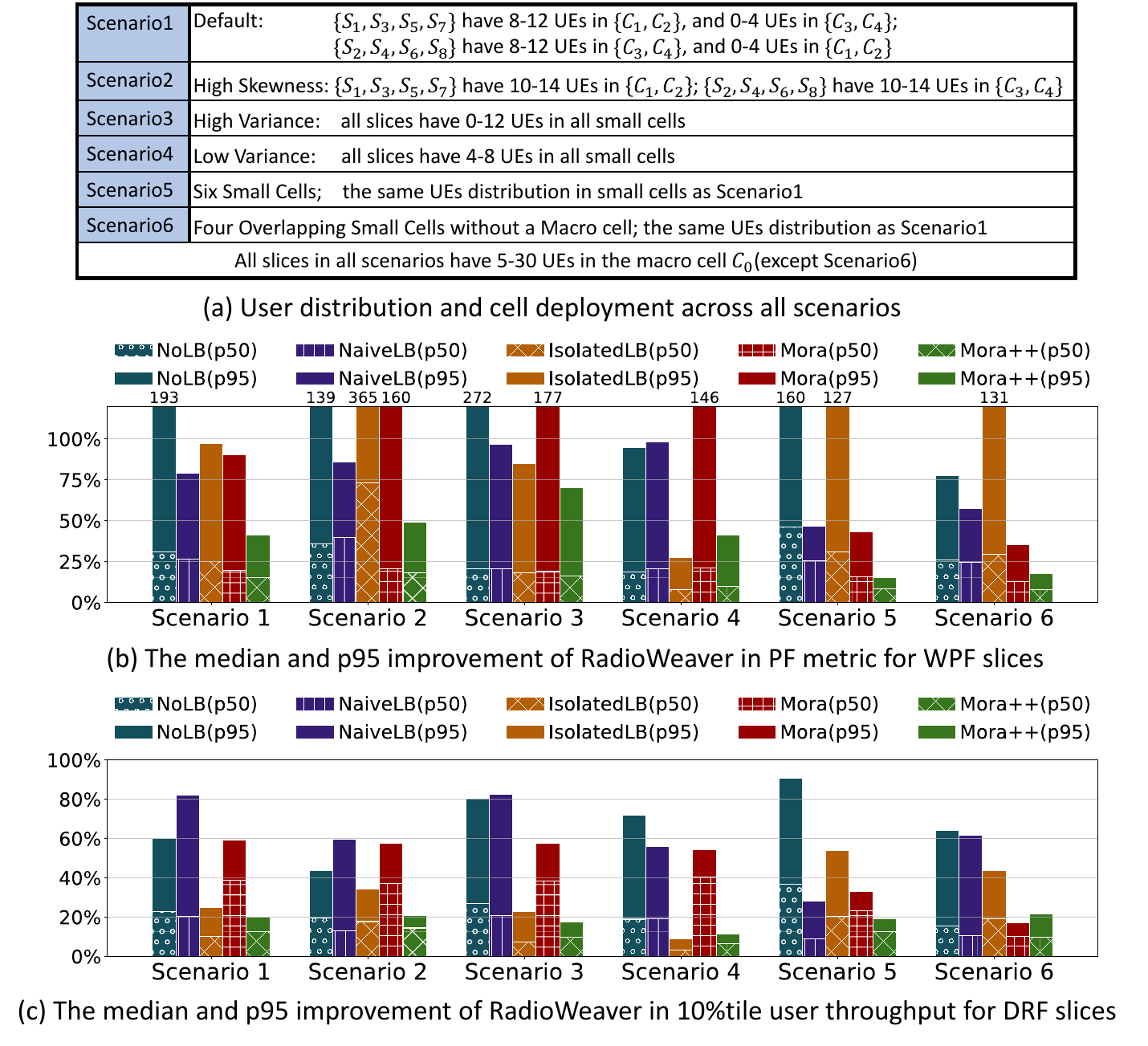}
    \vskip -0.1in
    \caption{The experiment settings and performance improvement of \oursys compared with other baselines in multiple scenarios.}
    \label{fig:multi-scenarios}
    \vskip -0.15in
\end{figure}

\paragraphb{1: Varying Geo-distribution and Number of Cells}
We now evaluate \oursys's performance compared to the baselines as we vary the geo-distribution of users and the number of cells. Fig\ref{fig:multi-scenarios}(a) elaborate the details of six scenarios we evaluated. Scenario1 is the default scenario as described in \S\ref{subsec:case1}. From Scenario 2 to Scenario 4, we change the geo-distribution of users in small cells (using the same cell deployment pattern as Scenario 1). Scenario 2 captures a more skewed geo-distribution (where some slices have no users in the coverage regions of some small cells). Scenario 4 has balanced geo-distribution of users for each slice across each small cell. Scenario 3 is similar, but exhibits higher variance across experiment runs. 
In Scenario 5, we increase the number of small cells to six. 
In Scenario 6, we remove the macro cell, and place four small cells with overlapping areas (Appendix\S\ref{subsec:overlap-smallcells} provides more details). The first four slices optimize for WPF, while the next four optimize for DRF (as described for Scenario 1).

Fig\ref{fig:multi-scenarios}(b) and (c) summarize the results for each of these scenarios, presenting the \oursys's performance improvement when compared to each baseline.
Fig\ref{fig:multi-scenarios}(b) shows the median and 95\%ile improvement in weighted PF metric with \oursys compared with other LB baselines, computed across the 100 WPF slices over 25 experiment runs. Fig\ref{fig:multi-scenarios}(c) shows the corresponding improvements in the 10\%ile user throughput across the 100 DRF slices. \oursys consistently outperforms other baselines with different user distribution and cell deployment. IsolatedLB's performance for the tail set of slices is particularly poor for Scenario 2 (with higher skew) due to lack of amenable users within individual slices for balancing load across cells. \oursys's slice-agnostic hand over of amenable users enables it to balance the load (based on TND criteria) even in highly skewed settings, leading to 365\% improvement for WPF slices (in tail case) and 60\% improvement for DRF slices over IsolatedLB.  
The closest baseline again is MORA++, but \oursys's performance is still 50\% higher in the tail case for WPF slices for Scenario 1-3 and about 20\% higher for DRF slices -- and that too with far fewer handovers (as highlighted earlier). 
The performance gain over IsolatedHO and MORA++ is lowest in Scenario4 due to the relatively more balanced geo-distribution.
The performance of NaiveLB and MORA are the worst in most cases due to the misaligned slice demand estimation.

\vskip 0.04in \paragraphb{2: Different Slice Quotas} Our scenarios so far considered equal global quotas across eight slices, but differing criteria across slices. We now evaluate a scenario with the same user distribution across slices as above -- all eight slices optimize for the same criteria (WPF, configured as described above), but have different global quotas. Specifically, the global quota of $S_1$ and $S_2$ (in terms of their share of overall RAN capacity) is 3$\times$ higher than the remaining six slices.
Fig\ref{fig:case3}(d) shows the CDF of weighted PF metric improvement in every slice. We continue to see strong improvements with \oursys.  In the median case, \oursys's performance improvement is 20\% higher than MORA++ and 24\% higher than IsolatedHO, while in the tail case it is 105\% and 141\%.

\vskip 0.04in \paragraphb{3: \oursys Design Parameters} We repeated our experiments in \S\ref{subsec:case1} using a tighter threshold for amenable users ($\alpha$ = 0.9 instead of default 0.8). It resulted in a slight decrease in performance (6\% lower improvement in 10\%ile throughput of DRF slices) when compared to default \oursys, by allowing slightly fewer handovers for load balancing. We also ran an experiment disabling \oursys's second phase of slice-specific handovers -- this also resulted in a small decrease in performance (4\% lower improvement in 10\%ile throughput), highlighting how the key performance benefits of \oursys over other baselines come from its handover of amenable users.    

\subsection{Deployment and Evaluation in Colosseum}
\label{subsec:colosseum}

We implement \oursys in SCOPE, and deploy it on the open-source Colosseum testbed~\cite{colosseum}. Due to limited support for multi-cell scenarios in Colosseum, our evaluation is restricted to the setting with each UE has the same channel quality across all cells. Our goal here is to evaluate \oursys's implementation, rather than the goodness of its algorithm compared to other baselines (that we thoroughly evaluate through our simulations).  Consequently, we only compare \oursys with NoLB and NaiveLB in this simplified setting. 

\begin{figure}[t!]
    \centering
    \includegraphics[width=0.48\textwidth]{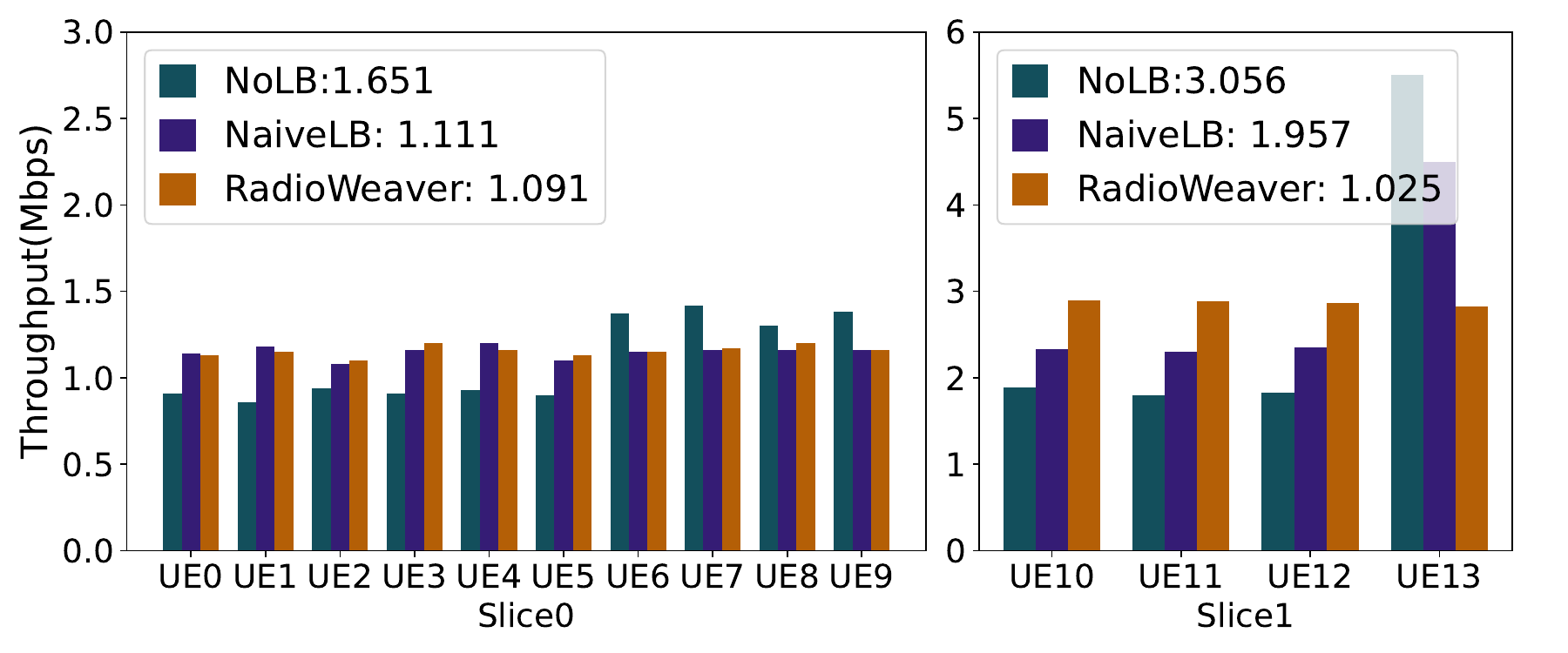}
    \vskip -0.15 in
    \caption{The average throughput of every UE in two slices. The legend shows the max-min fairness index(maximal user throughput divided by minimal user throughput) of NoLB, NaiveLB, and \oursys.}
    \vskip -0.15 in
    \label{fig:colos}
\end{figure}

We reserve two base stations with 20Mhz bandwidth and 14 users for our experiments. We have two slices with same global quotas: UE0-UE9 are assigned to $S_0$, and UE10-UE13 are assigned to $S_1$. We configure each slice to optimize for datarate fairness, and with equal weights across users.  
Assume initially 6 $S_0$ users (UE0-UE5) and 3 $S_1$ users (UE10-UE12) are attached to $C_1$, and 4 $S_0$ users (UE6-UE9) and 1 $S_1$ user (UE13) are attached to $C_2$ with NoLB. The RAN resources at each cell are split equally between the two slices, with users at $C_2$ enjoying higher throughput than users at $C_1$ (as shown in Fig\ref{fig:colos}). Suppose NaiveLB randomly transfers two UEs in $S_0$ from $C_1$ to $C_2$ to equalize the total number of UEs at each cell (7 each). This allows the RAN resources to be split between $S_0$ and $S_1$ in 40-60 ratio at $C_1$ and 60-40 ratio at $C_2$. $S_0$ achieves desired fairness across its UEs, but UE13 in $S_1$ (attached to $C_2$) enjoys 2$\times$ higher throughput than other UEs in $S_1$. \oursys appropriately transfers two $S_1$ UEs from $C_1$ to $C_2$ instead, achieving equal throughput across all users. 





\section{Conclusion}

With increasing density of cell deployments, a given user can have multiple options for its serving cell. The serving cell for each user must be carefully chosen such that the user achieves reasonably high channel quality from it, and the load on each cell is well balanced. This paper considers this problem in the context of slicing, a key 5G feature, where the network resources are divided among slices, with each slice optimizing its own performance criteria across its own group of users. Our system, \oursys, provides a new way to reason about global load balancing across users, that necessarily penetrates slice boundaries, but still respects the individual criteria, SLAs, and demand distributions of individual slices. 
\newpage
\bibliographystyle{abbrv} 
\begin{small}
\bibliography{paper}
\end{small}

\appendix
\clearpage
\appendix

\section{Measurement}
\label{sec:measure}
\begin{figure*}[t!]
    \centering
    \includegraphics[width=\textwidth]{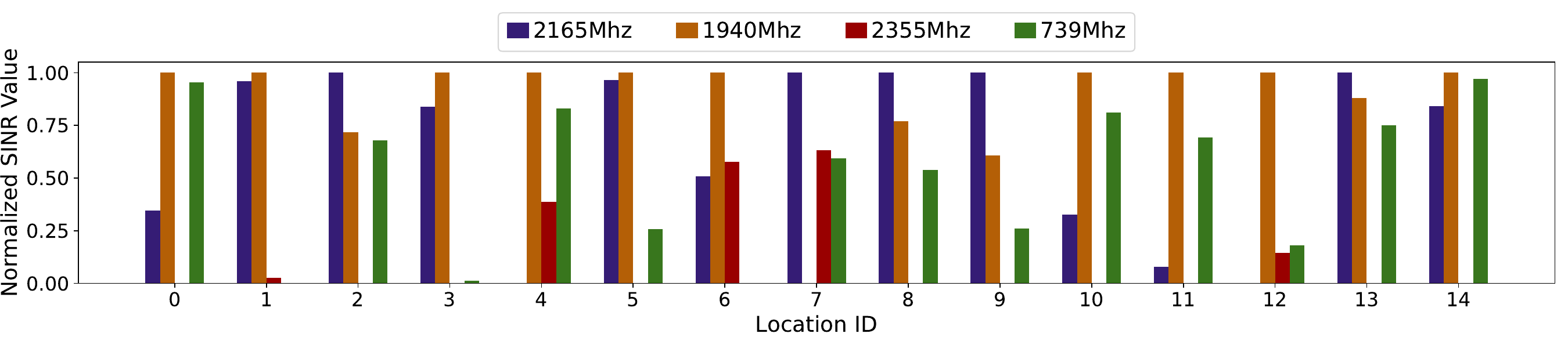}
    \vskip -0.1 in
    \caption{The normalized SINR values measured at 15 different locations for cellular operator A}
    \label{fig:att-measure}
    \vskip -0 in
\end{figure*}

\begin{figure*}[t!]
    \centering
    \includegraphics[width=\textwidth]{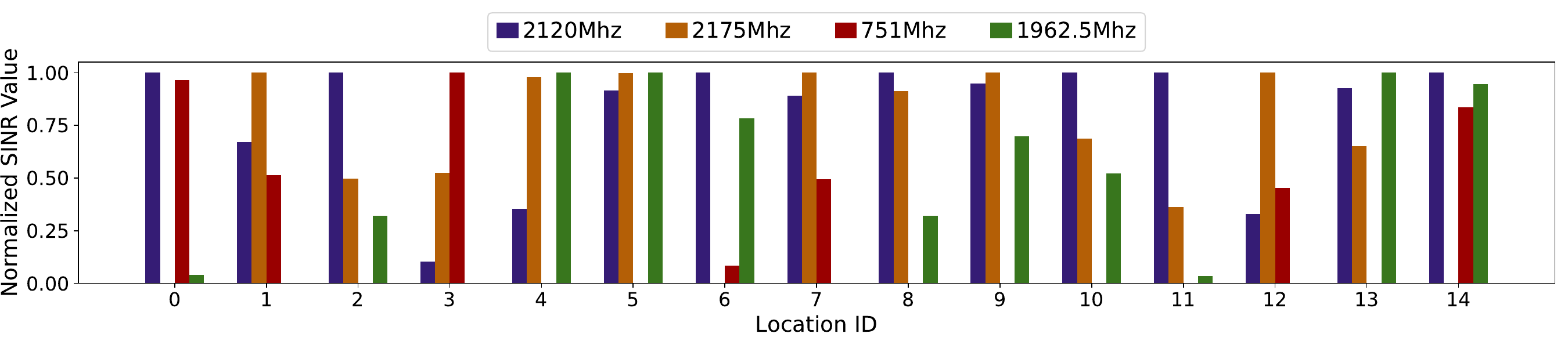}
    \vskip -0.1 in
    \caption{The normalized SINR values measured at 15 different locations for cellular operator B}
    \label{fig:verizon-measure}
    \vskip -0 in
\end{figure*}

We use NGScope\cite{ngscope} to collect the channel quality of cells operating in different bands from the same mobile network operator at different locations in a campus area. Fig\ref{fig:att-measure} shows the normalized SINR values of four cells from operator A measured at 15 different locations. At most locations, the user can experience similar high channel quality from two cells. However, the number of cells that can provide high channel quality could also be one or three. Fig\ref{fig:verizon-measure} shows the normalized SINR values of four cells from operator B, and it also shows the similar trend. In both figures, if the normalized SINR value is zero, it indicates the signal is so weak at that location that NGScope cannot even detect the cell.

\section{Proofs of Theorems}
\label{sec:proof-theorem}

We provide proofs of two lemmas and the theorem in \S\ref{subsec:theorem} in the following. We assume there are $K$ cells in total.

\subsection{Proof of Lemma 1}
\label{subsec:proof-lemma1}

Given a user distribution $\mathcal{U}$, now we prove that $P_i(\mathcal{U}, \mathcal{Q})$ is maximized for an allocation scheme $\mathcal{Q}$ where $\forall k, Q_{ik} = D_{ik} = d_{ik} \times Q_i$. For WPF slices, we have:

\begin{align*}
P_i(\mathcal{Q}) & = \sum_{k}\sum_{u_{ij} \in C_k} w_{ij}log(t_{ij}) \\
& = \sum_{k}\sum_{u_{ij} \in C_k} w_{ij}log(\frac{Q_{ik} w_{ij} }{\sum_{u_{ij} \in C_k}w_{ij}}e_{ij})
\end{align*}

Notice that $d_{ik} = \frac{\sum_{u_{ij} \in C_k} w_{ij}}{\sum_{u_{ij}} w_{ij}}$. Here, $\sum_{u_{ij}} w_{ij}$ is the sum of user's weight of all users in $S_i$ and is a constant. We use $N_i$ to annotate $\sum_{u_{ij}}w_{ij}$, so $\sum_{u_{ij} \in C_k} w_{ij} = d_{ik} N_i$.

\begin{align*}
P_i(\mathcal{Q}) & = \sum_{k}\sum_{u_{ij} \in C_k} w_{ij}log(\frac{Q_{ik} w_{ij} }{N_i d_{ik}}e_{ij}) \\
& = \sum_{k}\sum_{u_{ij} \in C_k} w_{ij}log(\frac{Q_{ik}}{d_{ik}}) + \sum_{k}\sum_{u_{ij} \in C_k} w_{ij}log(\frac{w_{ij}e_{ij}}{N_i}) \\
& = \sum_{k} d_{ik}N_i log(\frac{Q_{ik}}{d_{ik}}) + C_1
\end{align*}

To maximize $\sum_k d_{ik} log(\frac{Q_{ik}}{d_{ik}})$ with the constraint $\sum_k Q_{ik} = Q_i$, we use the method of Lagrange multipliers to find the maxima. The Lagrangian is $L(Q_{i1}, ..., Q_{iK}, \lambda) = \sum_k d_{ik} log(\frac{Q_{ik}}{d_{ik}}) + \lambda (\sum_k Q_{ik} - Q_i)$, and its local maxima must meet the following conditions:

\begin{align}
\begin{cases}
    \forall k, \frac{\partial L}{\partial Q_{ik}} & = 0 \\
    \frac{\partial L}{\partial \lambda} & = 0
\end{cases}
\label{condition1}
\end{align}

The only solution that meets condition\ref{condition1} is $\forall k, Q_{ik} = d_{ik} Q_i$. It's not hard to verify that this critical point is a local maximum point. Since $P_i(\mathcal{Q})$ is a strictly concave function, and the constraint is a hyperplane, the global maximum points can only be founded on the hyperplane. Therefore, $\forall k, Q_{ik} = d_{ik} Q_i$ is also the solution for the global maximum.

\begin{align*}
P_i(\mathcal{Q}) & = \sum_{k} d_{ik}N_i log(\frac{Q_{ik}}{d_{ik}}) + C_1 \\
& \leq \sum_{k} d_{ik}N_i log(Q_i) + C_1 \\
& = N_i log(Q_i) + C_1
\end{align*}

For WDRF slices, we want to maximize the minimal normalized user's throughput($\frac{t_{ij}}{w_{ij}}$) across all cells. Since all users of the same cell have the same normalized throughput, we use $T_k$ to annotate it in cell $C_k$. We have:

\begin{align*}
T_k = \frac{t_{ij}}{w_{ij}} &= \frac{w_{ij} / e_{ij}}{\sum_{u_{ij} \in C_k}w_{ij} / e_{ij}} \times Q_{ik} \times \frac{e_{ij}}{w_{ij}} \\
&= \frac{Q_{ik}}{\sum_{u_{ij} \in C_k}w_{ij} / e_{ij}}
\end{align*}

Recall that the demand ratio of a WDRF scheduling slice $S_i$ at $C_k$ is $d_{ik} = \frac{\sum_{u_{ij} \in C_k} w_{ij} / e_{ij}}{\sum_{u_{ij}} w_{ij} / e_{ij}}$, and the denominator here is a constant. We use $M_i$ to annotate $\sum_{u_{ij}} w_{ij} / e_{ij}$, so $\sum_{u_{ij} \in C_k} w_{ij} / e_{ij} = d_{ik} M_i$. Therefore, $T_k = \frac{Q_{ik}}{d_{ik}M_i}$.

\begin{align*}
P_i(\mathcal{Q}) & = min\{T_1, \cdots, T_K\} \\
& = min\{\frac{Q_{i1}}{d_{i1}M_i}, \cdots, \frac{Q_{iK}}{d_{iK}M_i}\}
\end{align*}

with constraints:
\begin{align}
\begin{cases}
    \sum_{k}Q_{ik} & = Q_i \\
    \sum_{k}d_{ik} & = 1
\end{cases}
\label{condition2}
\end{align}

It's not hard to find that the objective is maximized when $Q_{ik} = d_{ik} Q_i$ holds in every cell. The maximized minimal user throughput is $\frac{Q_i}{M_i}$ in slice $S_i$.

In conclusion, we have proven that $S_i$'s objective metric is less or equal to an upperbound, and the metric is equals to that upperbound if and only if the allocated quota equals the demand ratio times the cell's capacity in every cell.

\subsection{Proof of Lemma 2}
\label{subsec:proof-lemma2}

There are two constraints for any allocation scheme $\mathcal{Q}$: (i) The quota allocated to each slice $S_i$ summed across all cells must be equal to the slice's global quota, \ie $\sum_{k}Q_{ik} = Q_i$, and (ii) The sum of quota allocated to each slice at cell $C_k$ must be equal to the cell's capacity, \ie $\sum_{i}Q_{ik} = R_k$

We use $\mathcal{A}$ to refer the statement "$\mathcal{U}$ is fully complementary", and $\mathcal{B}$ to refer the statement "\emph{every} slice $S_i$ can achieve performance $P_i^{max}(\mathcal{U})$". We aim to prove that $\mathcal{A} \Leftrightarrow \mathcal{B}$, which requries showing both $\mathcal{A} \Rightarrow \mathcal{B}$ and $\mathcal{B} \Rightarrow \mathcal {A}$.

\noindent \textbf{Proof of $\mathcal{A} \Rightarrow \mathcal{B}$}:

Since $\mathcal{U}$ is fully complementary, it indicates that the TND at each cell equals to the cell's capacity, \ie $\forall k, \sum_i D_{ik} = R_k$. The allocation strategy is to assign $D_{ik}$ to $S_i$ in $C_k$, \ie $Q_{ik} = D_{ik} = d_{ik} Q_i$. This allocation meets both two constraints mentioned: 1. $\forall i, \sum_k Q_{ik} = Q_i$; 2. $\forall k, \sum_i Q_{ik} = R_k$.

\noindent \textbf{Proof of $\mathcal{B} \Rightarrow \mathcal{A}$}:

Assume $\mathcal{B}$ holds but $\mathcal{A}$ doesn't hold. If $\mathcal{U}$ is not fully complementary, there must be a cell $C_k$ in which $L_k = \sum_i D_{ik} > R_k$(if one cell is underloaded, there must be at least one cell which is overloaded). Therefore, it's impossible to allocate every slice the quota which equals its normalized demand in $C_k$. There must be a slice $S_i$ such that $Q_{ik} < D_{ik}$. Based on Lemma 1, $S_i$ cannot achieve performance $P_i^{max}(\mathcal{U})$, and this is contradictory to $\mathcal{B}$. As a result, if $\mathcal{B}$ holds, then $\mathcal{A}$ must also hold. \hfill \break

\subsection{Proof of Theorem 1}
\label{subsec:proof-theorem1}
Based on Lemma 2, if $\mathcal{U}^*$ is fully complementary, both slices that optimize for WPF and slices that optimize for DRF should achieve the upper bound calculated in Lemma 1.

For WPF slices, the upper bound is:

\begin{align*}
P_i(\mathcal{Q}) & = N_i log(Q_i) + C_1\\ 
& = N_i log(Q_i) + \sum_{k}\sum_{u_{ij} \in C_k} w_{ij}log(\frac{w_{ij}e_{ij}}{N_i})
\label{metric1}
\end{align*}

Here, $P_i(\mathcal{Q})$ increases strictly with the channel quality of every user($e_{ij}$) in $S_i$. Since $\mathcal{U}^*$ has optimal quality, every user should achieve the maximal effective datarate per RB. Therefore, every slice can achieve the globally maximum performance $\max_{{\mathcal{U}}} P^{max}_i(\mathcal{U})$.

For DRF slices, the upper bound is:
\[
P_i(\mathcal{Q})  = \frac{Q_i}{M_i}  = \frac{Q_i}{\sum_{u_{ij}} w_{ij} / e_{ij}}
\]

$P_i(\mathcal{Q})$ also increases strictly with the channel quality of every $e_{ij}$. Therefore, all DRF slices should also achieve the globally maximum performance if $\mathcal{U}^*$ has optimal quality.

\section{Auxiliary Evaluation Results}

\subsection{Multiple overlapping small cells}
\label{subsec:overlap-smallcells}
In the scenario of multiple overlapping small cells without a macro-cell, \oursys still applies the approach of two-step handovers over multiple rounds. In every round, \oursys iterates every overloaded small cell, and similarly hand over users from it to underloaded neighbor small cells. For a given overloaded cell($C_o$), \oursys moves all amenable users to each neighbor small cell until the load-driven stopping criteria is met in the first phase. If the load on $C_o$ is still higher than any neighbor cell($C_u$) after the first phase, \oursys transitions to the second phase and does slice-specific handovers from $C_o$ to $C_u$. The order in which \oursys sorts candidate UEs for handovers is the same as \S\ref{subsec:handover}.

\S\ref{subsec:vary} evaluated the performance of \oursys and other baselines in Scenario6, which has four homogeneous small cells with overlapping areas. In detail, the four small cells are placed 500 meters away from each other and operate in different bands. Fig\ref{fig:multi-scenarios}(b) and (c) shows that the median and tail slice-specific objective metrics of \oursys are consistently higher than that of other baselines. For the 10\%ile user throughput in DRF slices, \oursys outperforms MORA++ by 10\% at the median and 21.4\% at the 95th percentile. For weighted PF metric in WPF slices, \oursys outperforms MORA++ by 8\% at the median and 17.5\% at the 95th percentile.




\end{document}